\documentstyle[psfig]{mn}

\title{The extended star formation history of the star cluster NGC 2154 in the
Large Magellanic Cloud}

\author[Baume et al.]
{G. Baume$^{1}$,
 G. Carraro$^{2,3}$,
 E. Costa$^4$,
 R. A. M\'endez$^4$ and
 L. Girardi$^5$
\thanks{email: gbaume@fcaglp.fcaglp.unlp.edu.ar
(GB), giovanni.carraro@unipd.it (GC),
costa@das.uchile.cl (EC), rmendez@das.uchile cl (RAM),
leo.girardi@oapd.inaf.it (LG) }\\
$^1$Facultad de Ciencias Astron\'omicas y Geof\'{\i}sicas de la UNLP,
IALP-CONICET, Paseo del Bosque s/n, La Plata, Argentina \\
$^2$Dipartimento di Astronomia, Universit\`a di Padova, Vicolo
Osservatorio 2, I-35122, Padova, Italy\\
$^3$Andes Prize Fellow, Universidad de Chile and Yale University\\
$^4$Departamento de Astronom\'ia, Universidad de Chile, Casilla 36-D,
Santiago, Chile \\
$^4$Dipartimento di Astronomia, Universit\`a di Padova, Vicolo
Osservatorio 2, I-35122, Padova, Italy\\
$^5$INAF, Osservatorio Astronomico di Padova, Vicolo Osservatorio 5,
I-35122, Padova, Italy
}

\date{\it Submitted: Dec 2006}
\pubyear{2006}
\begin{document}
\maketitle
\title{The star cluster NGC~2154}

\begin{abstract}
The color-magnitude diagram (CMD) of the intermediate-age Large
Magellanic Cloud (LMC) star cluster NGC~2154 and its adjacent field,
has been analyzed using Padova stellar models to determine the
cluster's fundamental parameters and its Star Formation History (SFH).
Deep $BR$ CCD photometry, together with synthetic CMDs and Integrated
Luminosity Functions (ILFs), has allowed us to infer that the cluster
experienced an extended star formation period of about 1.2~Gyrs, which
began approximately 2.3~Gyr and ended 1.1~Gyr ago. The physical reality of
such a prolonged period of star formation is however questionable, and could
be the result of inadequacies in the stellar evolutionary tracks themselves.
A substantial fraction of binaries (70$\%$) seems to exist in NGC~2154.
\end{abstract}


\begin{figure*}
\centerline{\psfig{file=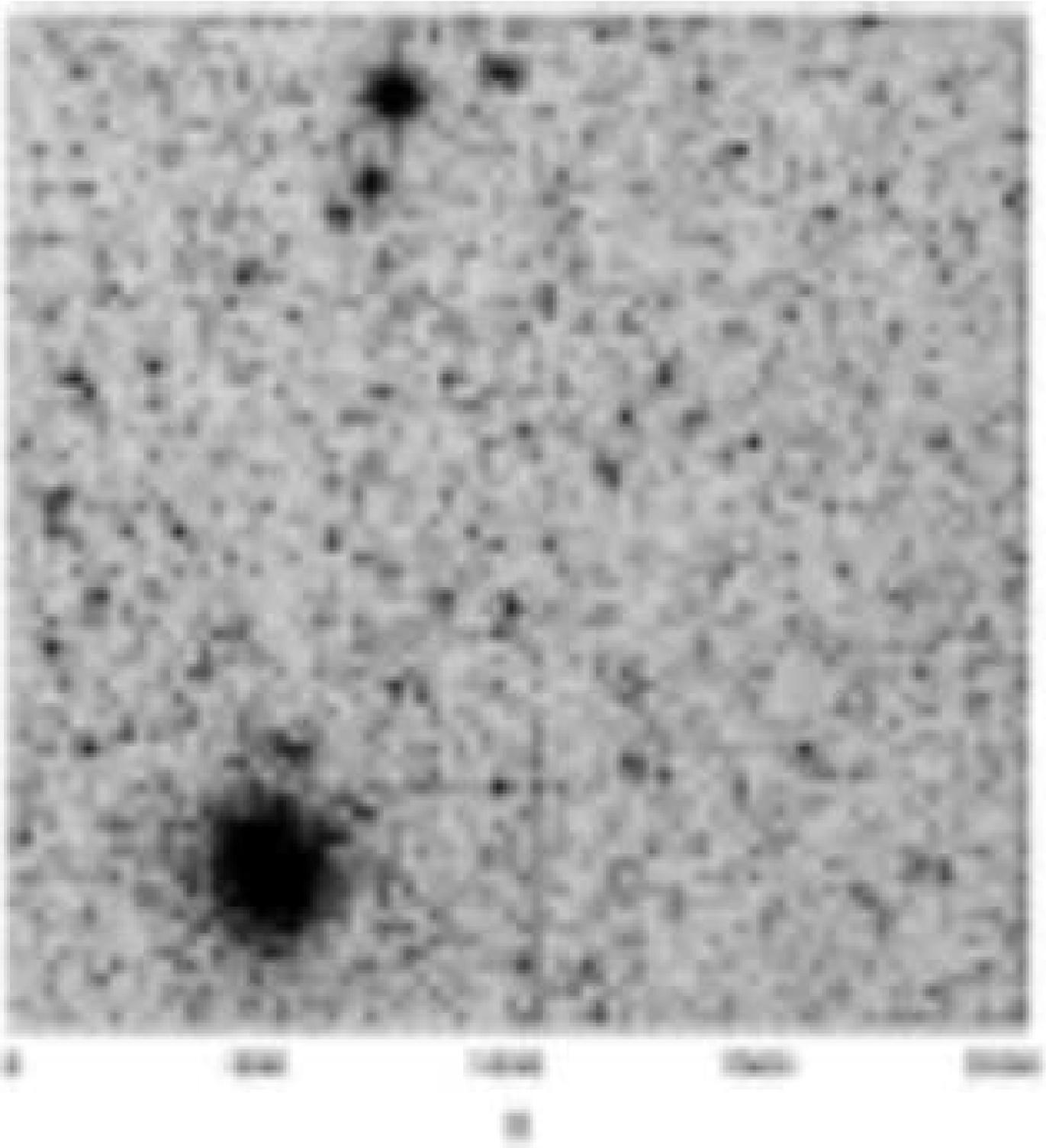}}
\caption{$R$ band image of NGC~2154 and the adjacent LMC stellar field.
North is up, East to the left. The size of the field is $8\farcm85
\times 8\farcm85$}
\label{fig:xy}
\end{figure*}

\section{Introduction}

The star cluster system of the Magellanic Clouds (MCs) differs
significantly from that of the Milky Way (and also from one another),
differences which are commonly attributed to a different chemical and
dynamical evolution.  Furthermore, MC clusters exhibit a broad range
of properties in contrast to our galaxy, thus representing a more
ample range of stellar populations than those represented by Galactic
clusters.  For the above reasons, MCs clusters have become a
challenging domain for stellar and galactic evolutionary models, and
are routinely used as an observational workbench to address these
issues (see e.g. Barmina et al. 2002, Bertelli et al. 2003, Woo et
al. 2003).  A very specific case is that of the intermediate-age,
metal-poor populations, which are conspicuous in the MC, yet rather
poorly represented in our galaxy.\\

One of many examples of intermediate-age clusters in the LMC is NGC~2154.
Although this cluster is morphologically globular, it is considered to be
of intermediate age (SMB type V; Searle et al. 1980).  Persson et al. (1983)
were able to constrain its age to the 1-3 Gyr range, but a detailed study
of its basic parameters (requiring high-precision deep photometry) was still
lacking.\\

Here we present deep CCD photometry, reaching B,R$\sim$25, of NGC 2154
and its adjacent LMC field, which has allowed for an unprecedented
study of the cluster, and a first-ever study of the SFH of this LMC
field. The present work is one result of a more comprehensive study of
the MCs, which includes the study of their SFH (Noel et al. 2006;
submitted to AJ), and the determination of their absolute proper
motions with respect to background QSOs (see e.g. Pedreros et
al. 2006). One of the LMC QSO fields selected for the proper-motion
work by chance included the neglected LMC cluster NGC~2154, which gave
us the possibility to observe this cluster -and its surrounding field-
routinely during our four-year (2001-4) campaign, with the
(additional) motivation of determining not only its fundamental
parameters, but also the SFH of the field (which will be the subject
of a forthcoming paper; Girardi et al. 2007).\\

The layout of the paper is as follows. Sect.~2 and 3 describe the
observations and the reduction strategy.  In Sect.~4 we study the
cluster structure and derive an estimate of its radius. Sect.~5 and 6
deal with the CMDs and describe the derivation of the cluster
fundamental parameters.  In Sect.~7 we summarize our conclusions.

\section{Observations}

$B(R)_{KC}$ images of the NGC~2154 region in the LMC were acquired with a 24$\mu$
pixels Tektronix 2048$\times$2048 detector attached to the Cassegrain focus of
the du~Pont 2.5-meter telescope (C100) at Las Campanas Observatory, Chile.
Gain and read noise were 3 e-/ADU and 7 e-, respectively.  This set-up provides
direct imaging over a field of $8\farcm85 \times 8\farcm85$
with a scale of $0\farcs259$/pix.
This relatively large field of view allowed
us to include the cluster and a good sampling of the LMC field in all frames.
The field covered by the observations is shown in Fig.~\ref{fig:xy}.\\

Details on the available frames and their corresponding exposure times are listed
in Table~1.
The B and R band-passes were selected in order to satisfy both the
needs of the SFH and astrometry programs (for this latter it was mandatory to
obtain R-band images).  Typical seeing was about $0\farcs9$.\\

All frames were pre-processed in a standard way using the
IRAF\footnote{IRAF is distributed by NOAO, which are operated by AURA
under cooperative agreement with the NSF.} package CCDRED. For this
purpose, zero exposures and sky flats were taken every night.

\begin{table}
\begin{center}
\caption {Log-book of observations for NGC~2154.}
\begin{tabular}{lccr@{$\times$}l}
\hline
\multicolumn{1}{c}{Date} & Airmass & Filter & \multicolumn{2}{c}{Exp. Time} \\
                         &         &        & [sec. & N]                    \\
\hline
15-10-01 &  1.28 & $R$ &  60 & 1 \\
         &  1.28 & $R$ & 300 & 3 \\
\hline
16-10-01 &  1.27 & $R$ & 400 & 3 \\
\hline
17-10-01 &  1.28 & $B$ & 400 & 1 \\
         &  1.27 & $R$ & 400 & 3 \\
\hline
18-10-01 &  1.28 & $R$ & 400 & 3 \\
\hline
08-10-02 &  1.51 & $R$ &  60 & 1 \\
         &  1.49 & $R$ & 400 & 9 \\
         &  1.30 & $B$ & 400 & 1 \\
         &  1.29 & $B$ & 600 & 4 \\
\hline
09-10-02 &  1.51 & $R$ &  60 & 2 \\
         &  1.31 & $B$ &  60 & 1 \\
         &  1.30 & $B$ & 600 & 1 \\
         &  1.28 & $R$ & 300 & 3 \\
         &  1.28 & $R$ & 400 & 1 \\
\hline
10-10-02 &  1.29 & $B$ &  60 & 1 \\
         &  1.29 & $B$ & 600 & 1 \\
         &  1.28 & $R$ & 400 & 1 \\
\hline
11-10-02 &  1.37 & $B$ &  60 & 1 \\
         &  1.32 & $B$ & 600 & 4 \\
         &  1.30 & $R$ & 400 & 2 \\
         &  1.28 & $R$ & 300 & 4 \\
\hline
12-10-02 &  1.37 & $B$ &  60 & 1 \\
         &  1.37 & $B$ & 600 & 3 \\
         &  1.28 & $R$ &  60 & 1 \\
         &  1.28 & $R$ & 300 & 3 \\
\hline
20-10-03 &  1.29 & $R$ &  60 & 1 \\
         &  1.29 & $R$ & 200 & 4 \\
         &  1.27 & $B$ & 800 & 3 \\
\hline
21-10-03 &  1.29 & $R$ &  60 & 1 \\
         &  1.29 & $B$ & 800 & 3 \\
         &  1.28 & $R$ & 300 & 4 \\
\hline
22-10-03 &  1.29 & $R$ & 600 & 1 \\
         &  1.27 & $R$ & 300 & 4 \\
\hline
24-10-03 &  1.31 & $R$ &  60 & 1 \\
         &  1.31 & $R$ & 600 & 2 \\
         &  1.29 & $R$ & 300 & 6 \\
\hline
04-11-04 &  1.30 & $R$ &  60 & 1 \\
         &  1.29 & $R$ & 600 & 6 \\
         &  1.28 & $B$ & 800 & 3 \\
\hline
05-11-04 &  1.31 & $R$ &   5 & 3 \\
         &  1.30 & $R$ &  30 & 3 \\
         &  1.30 & $R$ & 300 & 2 \\
         &  1.29 & $B$ &  10 & 4 \\
         &  1.28 & $B$ & 120 & 3 \\
         &  1.28 & $B$ & 600 & 2 \\
         &  1.27 & $R$ &  60 & 2 \\
         &  1.27 & $R$ & 300 & 5 \\
         &  1.27 & $B$ & 600 & 6 \\
\hline
\end{tabular}
\begin{minipage}{6cm}
\vspace{0.1cm}
$N$ indicates the number of frames obtained.
\end{minipage}
\end{center}
\end{table}

\begin{table} 
\tabcolsep 0.5truecm
\caption {Coefficients of the calibration equations}
\begin{tabular}{cc}
\hline
      & run 2001           \\
      & (four nights)      \\
\hline
$b_1$ & $+0.734 \pm 0.003$ \\
$b_2$ & $+0.218 \pm 0.002$ \\
$b_3$ & $-0.044 \pm 0.001$ \\
$r_1$ & $+0.434 \pm 0.003$ \\
$r_2$ & $+0.083 \pm 0.002$ \\
$r_3$ & $-0.001 \pm 0.001$ \\
\hline
\end{tabular}
\end{table}


\section{The Photometry}

\subsection{Standard star photometry}

Our instrumental photometric system was defined by the
use of the "Harris" UBVRI filter set, which constitutes the
default option at the C100 for broad-band photometry on
the standard Johnson-Kron-Cousins system.
On photometric nights, standard star areas from the
catalog of Landolt (1992) were observed multiple times
to determine the transformation equations relating our 
instrumental (b, r) magnitudes to the standard (B, R$_{KC}$) 
system. To determine atmospheric extinction optimally, a few 
of them were followed each night up to about 2.0 air-masses.
{\bf Besides, the standard star fields were selected to provide a wide color
coverage, being $-0.4 \leq (B-R) \leq 3.6$ (see Fig.~2)}.
Aperture photometry was carried out for all the standard stars 
using the IRAF DAOPHOT/PHOTCAL package.\\

To put our observations into the standard system, we used
transformation equations of the form:\\

\noindent
$b = B + b1 + b2 * X + b3 * (B - R)$\\
$r = R + r1 + r2 * X + r3 * (B - R)$\\

In these equations $b$, $r$ are the aperture magnitude already
normalized to 1 sec, and $X$ is the airmass. We did not include 
second-order color terms because they turned out to
be negligible in comparison to their uncertainties. The values 
of the transformation coefficients are listed in Table 2.
The night-to-night variation of the coefficients turned out to
be very small ($\sim$ 0.001), so we adopted the average values
over all nights. {\bf The residuals resulting from
the fits are showed in Fig.~2,  and the global rms of the 
calibration was 0.011 mag for both B and R filters.}

\subsection{Cluster and LMC field photometry}

Here we follow the procedure outlined in Baume et al. (2004).
We first averaged images taken the same night, and with
the same exposure time and filter, in order to get a higher
signal-to-noise ratio for the faint stars, and also to clean 
the images from cosmic rays. Then, instrumental magnitudes
and (X,Y) centroids of all stars in each frame were derived
by means of profile-fitting photometry with the DAOPHOT
package, using the Point Spread Function (PSF) method
(Stetson 1987); {\bf and all instrumental magnitudes from 
different nights were combined and carried to same system
(2001 reference) by using DAOMASTER (Stetson 1992). Stellar
magnitudes in the standard system were obtained then by
using the transformations indicated in section 3.1.}\\
This resulted in a photometric catalog  consisting of
about 20000 stars\\

\noindent
In Fig. 3 we present the photometric errors trends (from
DAOPHOT and DAOMASTER) as a function of the B magnitude. 
Down to $B = 22$, both the $B$ band and $R$ band errors remain 
lower than 0.025 mag. From Fig. 2 it can be easily seen that 
the main source of error originates in the B-band observations. 
It is this filter that determines the deepness of our 
photometry, a point that is relevant to the completeness 
analysis (see Sect. 3.3).

\subsection{Photometric completeness}

For all comparisons described in the next sections, completeness 
of the observed star counts is a relevant issue. Completeness 
corrections were determined by means of artificial-star experiments 
on our data (see Carraro et al. 2005). {\bf Basically, we created 
several artificial images by adding in random positions a total of 
40000 artificial stars to our true images. They were distributed with 
a uniform probability distribution with the same color and luminosity 
of the  real sample} . In order to avoid the creation of 
overcrowding, in each experiment we added the equivalent to only 15%
of the original number of stars.

Given that in general the $B$ band images are shallower than those in 
the $R$ band, we have adopted as completeness factor that estimated 
for $B$. This factor is defined as the ratio between the number of 
artificial stars recovered and the number of artificial stars added. 
Computed values of the completeness factor for different B magnitude 
bins are listed in Table 3, both for the cluster (r < 400 pix), and 
for a representative comparison field (see Sect. 6.1). {\bf It must 
be noticed that due to the inherent nature of a very compact star 
cluster (see Figs. 1 and 4), more than half of the stars would occupy 
less than half of the volume. As a consequence,  the completeness fractions
for the cluster stars are likely to be a bit overestimated. On the 
other hand, the use of a lower radius for the cluster region 
would have the disadvantage to imply  larger 
uncertainties in the completeness  factors.}


\begin{table}
\tabcolsep 0.5truecm
\caption {Completeness analysis results in the $B$ band.}
\begin{tabular}{ccc}
\hline
$\Delta B$ & NGC~2154    & Field \\
           & (r$<$400pix) &       \\
\hline
19.5-20.0  & 100.0\% & 100.0\% \\
20.0-20.5  &  92.7\% & 100.0\% \\
20.5-21.0  &  75.1\% & 100.0\% \\
21.0-21.5  &  57.4\% & 100.0\% \\
21.5-22.0  &  56.8\% & 100.0\% \\
22.0-22.5  &  56.1\% & 100.0\% \\
22.5-23.0  &  55.0\% & 100.0\% \\
23.0-23.5  &  53.9\% &  82.7\% \\
23.5-24.0  &  41.7\% &  61.1\% \\
24.0-24.5  &  29.5\% &  62.0\% \\
24.5-25.0  &  32.4\% &  65.6\% \\
25.0-25.5  &  35.2\% &  82.5\% \\
25.5-26.0  &  36.3\% &  49.6\% \\
\hline
\end{tabular}
\end{table}
\begin{figure}
\begin{center}
\centerline{\psfig{file=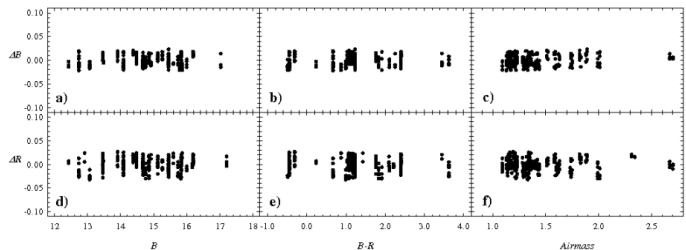,width=9cm}}
\caption{Trend of residuals of the standard stars calibration.}
\label{fig:err}
\end{center}
\end{figure}

\begin{figure}
\begin{center}
\centerline{\psfig{file=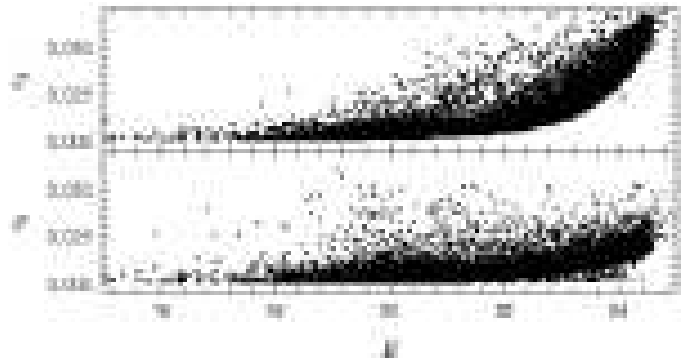,width=9cm}}
\caption{Photometric errors in $B$ and $R$ given by DAOPHOT and
DAOMASTER, as a function of the $B$ magnitude.}
\label{fig:err}
\end{center}
\end{figure}

\section{Star counts}

\subsection{Cluster radius}

In order to study the cluster structure, as a first step we estimated
the position of the cluster center by determining the highest peak in
the stellar density, which was done by visual inspection of the our
images.  This peak was found at: $X = 515$; $Y = 1725$, which
corresponds to $\alpha_{2000} = 5^{h}57^{m}38^{s}.2$; $\delta_{2000} =
-67^{o}15^{\prime}40^{\prime\prime}.7$, coordinates which are similar
to those given in the $SIMBAD$ database.\\

\begin{figure}
\begin{center}
\centerline{\psfig{file=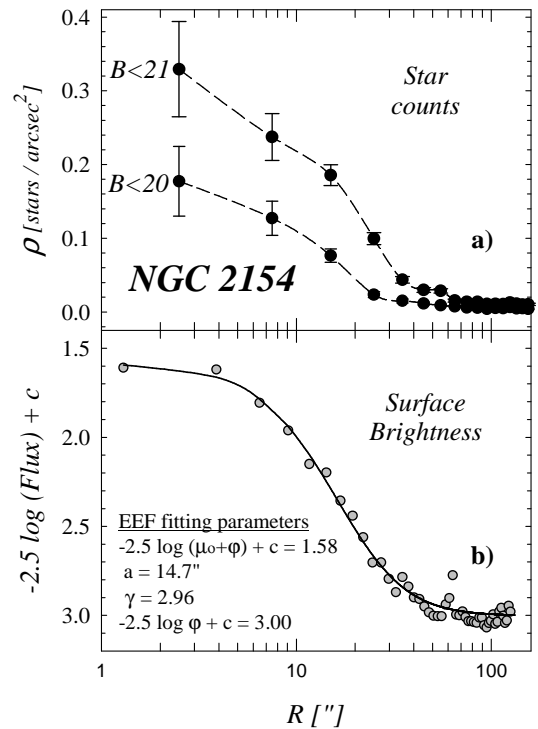,width=7cm}}
\caption{{\bf a)} Radial density profiles for NGC~2154. Numbers indicate the
limit $B$ magnitudes for each case. {\bf b)} Radial flux profile for NGC~2154
(grey circles) and the EFF model fit (solid curve) together with the computed
fitted parameters ($c$ is an arbitrary constant).}
\label{fig:rad}
\end{center}
\end{figure}

The next step was to compute the cluster's size, which was done
constructing radial profiles by two methods: the radial stellar
density profile and the the radial flux profile methods:\\

a) In the first, stars are counted in a number of successive rings,
$30''$ wide, concentric around the adopted cluster center, and then
divided by their respective areas. Because our data does not permit
complete annuli beyond 515 pix, we have assumed that the measurable
annuli portions are still representative of the field stellar
populations around the cluster.  The density profiles obtained, down
to two different $B$ limit magnitude limits (20 and 21) are shown in
Fig.~\ref{fig:rad}a.

b) In the second method, the flux ($-2.5~log~[ADUs/area]$) within
concentric annuli 10 pixels wide ($2\farcs59$) is measured directly
over the ($B$ band) cluster image. The resulting profile is presented
in Fig.~\ref{fig:rad}b.  A measure of the cluster's radius is obtained
by fitting a model from Elson, Fall and Freeman (EFF; see Elson et
al. 1987), appropriate for LMC clusters (Mackey \& Gilmore 2003).  The
expression used was:\\

\noindent $\mu = \mu_0 [1 + (r/a)^2]^{-\gamma/2} + \phi$ \\

\noindent where $r$ is the distance from the adopted cluster center,
$\mu_o$ the central surface brightness, $a$ is a measure of the core
radius, $\gamma$ the power law slope at large radii, and, finally,
$\phi$ is the field surface brightness.  The computed parameters are
given in Fig.~\ref{fig:rad}b.

\subsection{Observed Luminosity Function}

\begin{figure}
\begin{center}
\centerline{\psfig{file=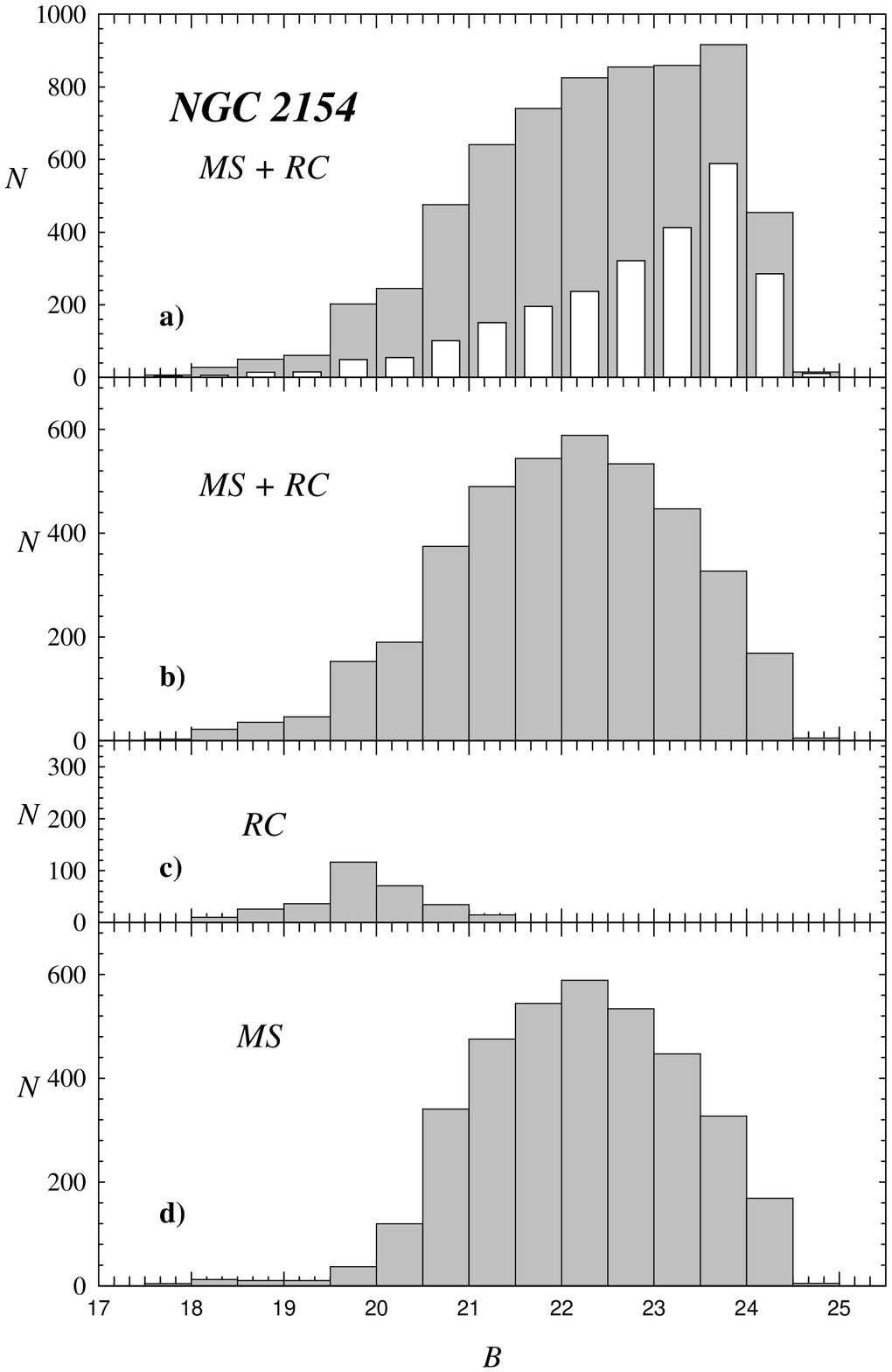,width=7cm}}
\caption{{\bf a)} LFs corrected by completeness factors. The grey
histogram corresponds to the 'cluster area', whereas the white
histogram to the 'field area'. {\bf b)} LF corrected by field
contamination. {\bf c)} and {\bf d)} Idem as panel {\bf b)} but only
for RC and MS regions respectively. (see Sect.~4.2).}
\label{fig:lf1}
\end{center}
\end{figure}

We have constructed {\it B}-band Luminosity Functions (LFs) of the
{\it cluster region} ($r < 400$~pix) and of a similarly size {\it
field region} (centered at $X = 1500$, $Y = 500$). The counts
presented were corrected using the completeness factors given in
Table~3 (see Sect. 3.3). The results are plotted in
Fig.~\ref{fig:lf1}a.\\ In Figure~\ref{fig:lf1}b we present the pure
cluster LF obtained by subtracting the {\it field region} LF from the
{\it cluster region} LF.\\

Finally, we separated the data (both in the {\it cluster region} and
in the {\it field region}) in two sets in order to isolate the red
clump (RC) stars: those with $B < 21.5$ and $B-R > 1.0$, from the rest
of the data (the main sequence region, MS). The resulting, field
subtracted, LFs for the RC and MS regions are given in
Fig.~\ref{fig:lf1}c and Fig.~\ref{fig:lf1}d, respectively.

\section{Color-Magnitude Diagrams}

In Fig.~6 we present the CMD of {\it all} stars measured in the
{\it complete} field shown in Fig.~1.  In Fig.~7 we present the CMD
of the region centered at X=515, Y=1725 and having $r \leq 400$ pix
(the {\it cluster region}, see Sect.~4).  Although well centered in the
cluster, this CMD is clearly contaminated by LMC field stars (it is
interesting to note that, as it is, this CMD closely resembles that of
the intermediate-age LMC cluster NGC~2173 studied by Gallart et al.
(2003)).\\

\begin{figure}
\centerline{\psfig{file=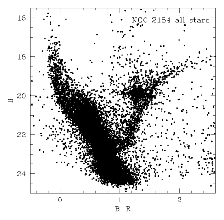,width=\columnwidth}}
\caption{NGC 2154 CMD for all the measured stars.}
\end{figure}

\begin{figure}
\centerline{\psfig{file=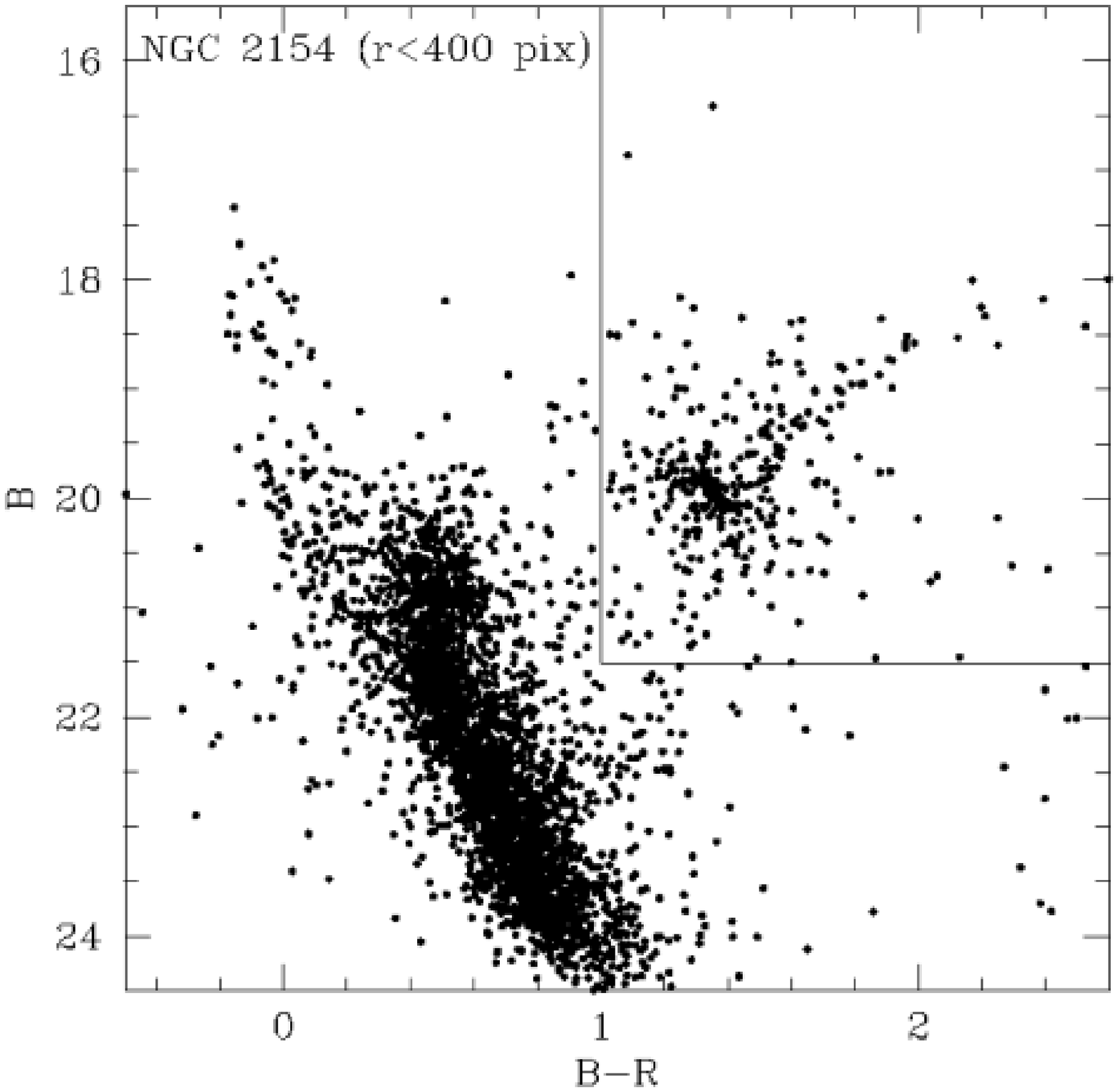,width=\columnwidth}}
\caption{NGC 2154 CMD for the stars lying within
400 pixel from the cluster center. The solid line encloses
the area where we looked for evolved stars.}
\end{figure}

To obtain a cleaner CMD, we applied the statistical decontamination
method described in Vallenari et al. (1992) and Gallart et al. (2003).
In this procedure, a statistical subtraction of field stars is carried out
making a star-by-star comparison between selected reference field regions
and the {\it cluster region}.  Briefly, for any given star in the reference
field regions we look for the most similar (in color and magnitude) star
in the {\it cluster region}, and remove it from the cluster's CMD.  It
should be noted that the procedure takes into account the different
completeness level of the cluster and the field.\\

For the above purpose, we selected three reference field regions having
the same area of the {\it cluster region}, which we call (see Fig.~8)
Field$\#1$ (centered at X= 1500,Y=500), Field$\#2$ (centered at
X=1500,Y=1500) and Field$\#3$ (centered at X=500,Y=500). They were
chosen at proper distances from the cluster center, in order to avoid
the presence of cluster stars in them. The results are shown in Fig.~8.
The three upper panels show the CMDs of the reference fields, whereas the
middle and lower panels show the corresponding CMDs of the subtracted stars,
and the corresponding clean cluster CMDs, respectively.\\

\begin{figure*}
\centerline{\psfig{file=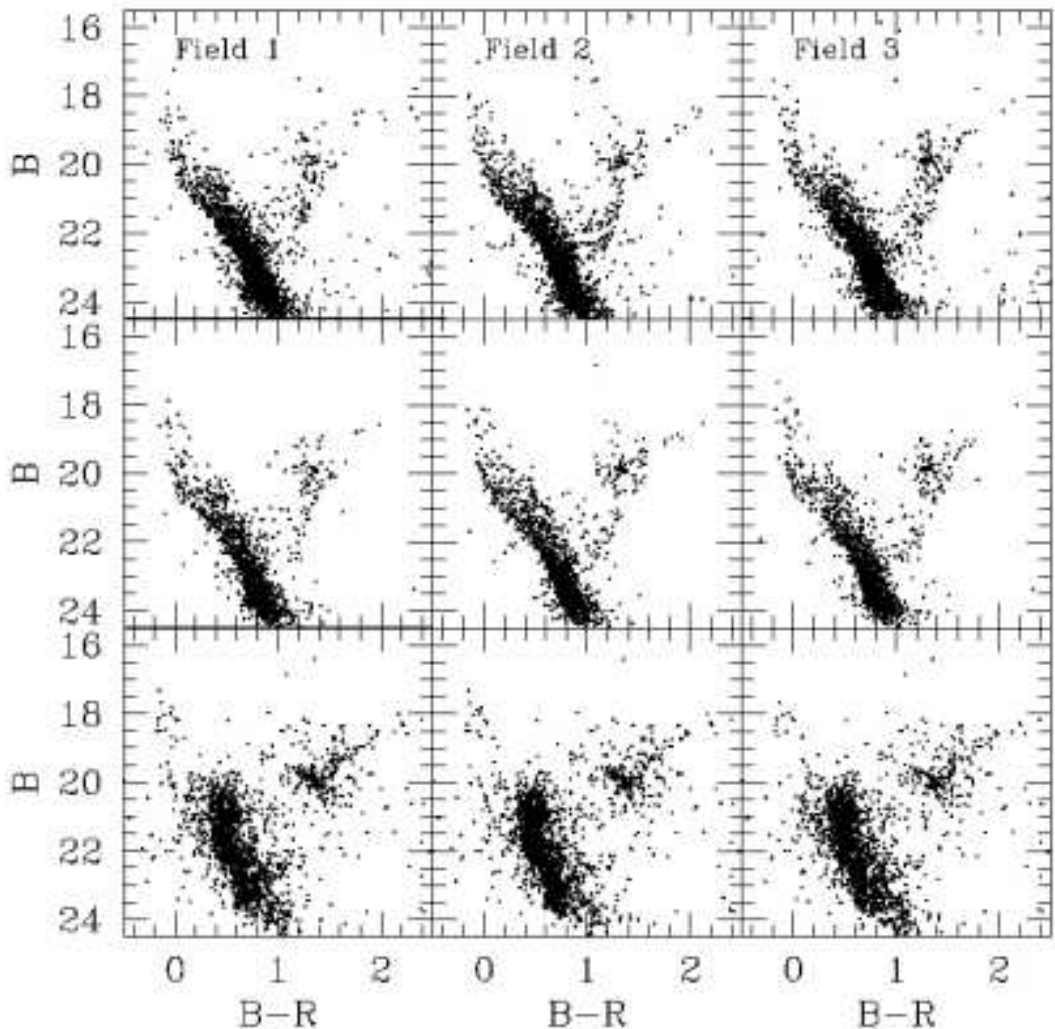,width=15cm,height=15cm}}
\caption{Field star decontamination procedure. The upper panel shows
the CMDs of three different field regions.  The middle panel shows the
corresponding CMDs of the subtracted stars. In the lower panel the
corresponding clean cluster CMDs are shown. See text for details.}
\end{figure*}

Probably due to peculiarities in the distribution of field stars
across the cluster region, a perfect decontamination was however not
possible.  Inspection of Fig.~8 shows that clean cluster regions
present several groups of stars that are more numerous than in the
reference field regions, despite the fact that all regions have the
same area. In the clean CMDs some field stars still remain above the
MS, to the left of the MS and everywhere in the red clump region. This
effect surely results from the statistically low number of stars in
those regions of the CMD.\\

Nonetheless, the procedure was very effective and helped us to improve
the shape of the turnoff region (TO), and to remove several field red
giant Branch (RGB) and horizontal branch stars.  A careful
inspection of Fig.~8 led us to adopt as the clean NGC 2154 CMD, that
obtained using Field $\#3$ for the statistical decontamination (lower
right panel). In this case, the subtracted field is the closest to the
corresponding original field, and the number of field stars still in
the cluster region is significantly lower than in the other two
cases.\\

\section{Comparison with stellar models}

In this section we derive estimates of NGC~2154's fundamental
parameters by comparing its CMD with theoretical stellar evolutionary
models from the Padova library of stellar tracks and isochrones
(Girardi et al. 2000).  These models have already been used in the
past to study LMC star clusters with satisfactory results
(e.g. Barmina et al. 2002, Bertelli et al. 2003). We first summarize
previous work on NGC~2154 and what we know from the literature of its
basic parameters.

\subsection{Metallicity}

Bica et al. (1986) used H$\beta$ and G-band photometry to estimate
that the metallicity of NGC~2154 is $Z=0.006$.  NGC~2154 was later
observed by Olszewski et al. (1991) as part of a spectroscopic survey
of giant stars in LMC star clusters, who derived a metallicity
measuring the pseudo-equivalent width of three Ca lines.  Two stars
were measured with this purpose, giving an estimate of the cluster's
metallicity of [Fe/H]=-0.56$\pm$0.20.  Adopting the Carraro et
al. (1999) relation, this value corresponds to Z = 0.005.  We shall
adopt this estimate of the metallicity throughout this work.

\subsection{Reddening and Distance modulus}

While the distance modulus to the LMC is known with reasonable
precision (Westerlund 1997, $(V_o-M_V)=18.5$), no estimates of the
reddening in the direction of NGC~2154 are available. To complicate
matters, the reddening across the galaxy is known to be highly
variable (Oestreicher \& Schmidt-Kaler 1996; Zaritsky et al. 2004).
As for the Galactic extinction law, here we shall use $R_V$=3.1 (Rieke
\& Lebofsky 1985).

\subsection{Isochrone fitting}

In Fig.~9 we have superposed three Z=0.005 ([Fe/H]= -0.60)
isochrones, taken from the Girardi et al. (2000) database, to our
adopted clean NGC 2154 CMD.  It should be noted that this metallicity
is not directly available, so it has been interpolated from different
metallicity datasets.  We have selected these isochrones because they
provide a good fit to the MS and the TO region, and also to the
magnitude and color of the RC.  They have been shifted by the
reddening and distance modulus indicated in the figure. The values
presented for these parameters imply a corrected distance modulus of
$(B_o-M_B)$ in the range 18.45 to 18.50, in good agreement with the
widely adopted distance modulus given by Westerlund (1997).\\

\begin{figure}
\centerline{\psfig{file=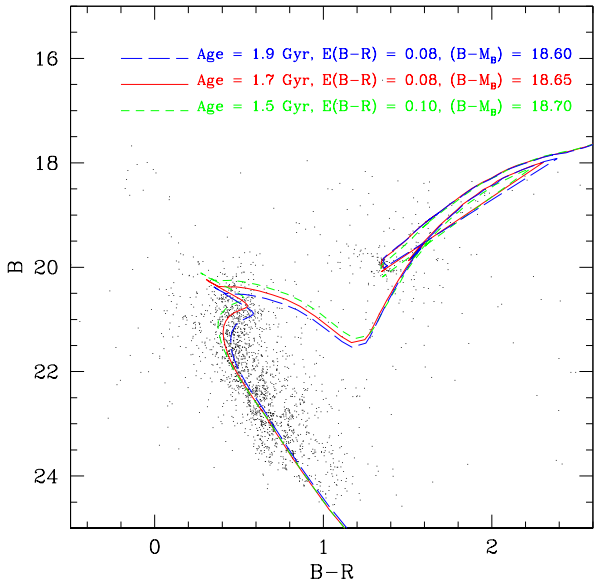,width=\columnwidth}}
\caption{CMD of NGC~2154 decontaminated from field stars (see see Sect.~5).
The three $Z=0.005$ isochrones superposed are from the Girardi et al. (2000)
database (see Sect.~6.3).}
\end{figure}

Quick examination of Fig.~9 shows that:

\begin{itemize}
\item  The MS is significantly wide. This may be due to a combination of
observational errors, binarity, differential reddening, and age spread;
\item  The RGB clump is wide in color, and slightly tilted. This again can be
ascribed to differential reddening and binarity;
\item  The stars above the TO, and outside the MS edges are mostly interlopers
belonging to the LMC field in front of the cluster;
\item  On the other hand, the reddest field stars are probably members of the
Milky Way Halo.
\end{itemize}

In the next Section we test some of these interpretations by means of
synthetic CMDs.

\subsection{Synthetic Color-Magnitude Diagrams and Model Luminosity Functions}

Synthetic CMDs have been generated using the TRILEGAL code described
in Girardi et al. (2005).  The detailed procedure is outlined in
Carraro et al. (2002). Using typical values derived from our
observations (see Sect.~3), we have also simulated the photometric
errors as a function of $B$ and $R$ magnitude.\\

Using our results of the isochrone fitting, we generated a synthetic CMD
for a cluster which underwent an instantaneous burst of star formation 1.7
Gyr ago, and which has a population of 290 evolved stars. This number of
evolved stars was derived from the luminosity function discussed in
Sect.~4; and it is somewhat uncertain due to (1) possible errors while
removing the contamination by the field, and (2) Poisson
statistics. If we neglect for the moment the uncertainties in the
field contamination, and assume the Kroupa (2001) Inital mass Function
(IMF) corrected for binaries, this number of evolved stars implies a
cluster mass of $33.3\pm2.0\times10^3$~$M_{\odot}$.\\

Our initial CMD simulation includes a fraction f=30\% of
detached binaries, and assumes that their mass ratio is uniformly
distributed between 0.7 and 1.0. This assumption is in agreement with
the observational data for the LMC clusters NGC~1818 (Elson et
al. 1998), NGC~1866 (Barmina et al. 2002), and NGC~2173 (Bertelli et
al. 2003).  It is worth recalling that the photometry of a binary with
mass ratio smaller than 0.7 is almost indistinguishable from its
primary alone, so that extending the interval of simulated mass ratios
would not change the results. \\

The results for this initial choice of parameters are shown in
Fig.~10. The left panel shows the observational, decontaminated, CMD of
NGC 2154 as derived in the
previous section, and the right panel shows the result of our best
simulation; chosen among many synthetic CMDs generated with the same
input parameters but varying the random seed (Bertelli et al. 2003).
The top panel presents the observational (see Sect.~4.2) ILF
({\bf Integrated Luminosity Function}, solid line)
together with the corresponding best fitting model ILF (dashed line).\\

In the simulations, special attention was given to the shape of the TO region,
and to the color and magnitude of the RGB clump. Because of the complicated
structure of the TO, which is broadened by the presence of binaries and the
extended star formation period, for the derivation of the cluster's distance
and reddening we have used mainly the RGB clump. As can be readily seen, the
mean color and magnitude of the clump has been well reproduced,
allowing us to infer a reddening $E(B-R) = 0.09 \pm 0.02$ ($E(B-V) = 0.057$),
and distance modulus $(B-M_B)_0 = (V-M_V)_0 = 18.48 \pm 0.10$. Both these
values are well within the widely accepted estimates for the LMC.\\

The red clump stars, however, seem to present a more elongated, tilted,
distribution in the CMD, than in the initial simulation. This kind of
structure might be caused by an increased fraction of binaries, by
differential reddening, or by an age spread inside the cluster. All of
these effects would affect not only the CMD, but also its integrated
LF, and in different ways.\\

We have investigated these possible effects by running additional
simulations in which we varied the following quantities:

\begin{itemize}

\item
The fraction of binaries $f$, from 0 (no binaries) to 1 (each star
drawn from the IMF is the primary of a system with mass ratio between
0.7 and 1) at steps of 0.1. Because the binaries are added with equal
probability to both the MS and red giant part of the CMD, $f$
affects the predicted dwarf/giant ratio, and hence the ILF.
\item
The $1\sigma$ dispersion in reddening, between 0 and a maximum of 0.09
in $E(B-R)$. Of course, negative values of reddening are not allowed,
and become 0. This causes a maximum broadening of about 0.2 mag of the
red clump in $B-R$, without affecting the overall shape of the ILF.
\item
The duration of the star formation episode that generated the cluster,
from 0 to $2.0$~Gyr, with steps of about 0.2~Gyr, centered at an age
of 1.75~Gyr. This spread has a modest effect in the red clump
luminosity, but affects the turn-off region of the ILF.
\end{itemize}
For each model in the grid, we compute the reduced
$\chi^2$, defined as the mean squared difference between model and
observations, in the magnitude interval $16<B<21.5$. Fainter stars are
not included both because (1) below the turn-off region the ILF
becomes very sensitive to the IMF, which is not known with enough
accuracy; and (2) at those magnitudes the completeness corrections
become large and consequently more uncertain.\\

The model with the lowest value of $\chi^2$ (=347) has $f=0.7$,
$\sigma E(B-R)=0.09$~mag, and a SFH duration of 1.2~Gyr. The
corresponding best fitting synthetic CMD and model ILF are presented
in Fig.~11. The top panel presents the observational (see Sect.~4.2)
ILF (solid line) together with the model ILF (dashed line). This theoretical
ILF therefore represents the same population, with the same binary fraction,
same star formation history, and photometric errors as the synthetic CMD,
{\em and with as many red giants as in the observations}.  It can be
readily noticed that they  agree reasonably well down to $B=23.5$,
magnitude level at which the incompleteness corrections are still smaller
than 50$\%$.  It should be noticed however, that the differences between
them are statistically significant, i.e. larger than the 1$\sigma$ error
bars (67\% confidence level) determined from the Poisson statistics.\\

We would like to note that a solution of similar quality, with $\chi^2=549$,
is found for $f=0.4$, $\sigma
E(B-R)=0.0$~mag, and a SFH duration of 2~Gyr. This second solution has
reasonable values of $f$ (well in agreement with estimates for other
LMC clusters) and $\sigma E(B-R)$.  Despite the good quality of the
ILF fit as measured by the $\chi^2$, a 2~Gyr duration for the star
formation in NGC~2154 seems unacceptably long.\\

\section{Conclusions}

We have presented and discussed deep $BR$ photometry of the
intermediate-age open cluster NGC~2154 in the LMC. By using
theoretical tools, namely isochrones, synthetic CMDs, and an ILF, we
obtained estimates of the cluster fundamental parameters. The distance
and reddening found fall within the commonly accepted values for the
LMC.\\
\noindent
Very interestingly, we found that the cluster CMD, and in particular
the ILF, can be properly interpreted only allowing for a extended
period of star formation (1.2~Gyr). Another, less extreme, example of
extended period of cluster formation, is that of the almost coeval
cluster NGC~2173 in the LMC, with 0.3~Gyr (Bertelli et al. 2003).
Whether the presence of such extended period of star
formation is physically possible, is to be investigated in future
works. From our side, it is mandatory to mention that the detection of
such a prolonged star formation period could well be caused by
inadequacies in the stellar evolutionary tracks themselves. The most
obvious among these inadequacies is the uncertain efficiency of core
convective overshooting in MS stars of masses between 1 and
2~$M_\odot$ (see e.g. Chiosi 2006).\\

\noindent
{\bf An interesting point raised by the referee is whether the blue
stars which concur to enlarge the MS and thicken the TO
producing the extended star
formation effect we find, might indeed be Blue Stragglers Stars (BSSs). 
These stars are ubiquitous,
and they are routinely  found  in Dwarf Galaxies,
Galactic star clusters and the general Galactic field (de Marchi et
al. 2006). 
Unfortunately, a comprehensive search and analysis of BSSs in the LMC clusters
is still unavailable.\\
While there is no consensus yet 
on the mechanism which produces these stars, their position in the CMD and
their population are reasonably well understood. They occupy a strip
along the extension of the MS above the TO point, 
but tend to be bluer than the standard binary star
sequence (which runs parallel to the MS).
Besides, 
they occupy a region of the CMD where
young stars of the general stellar field toward a star cluster or a
dwarf galaxy are found. This complicates their detection and demand
more effective membership criteria.\\
In our CMD, after the cleaning procedure, most of blue stars are actually more
compatible with field stars, and the shape of the TO point does not
seem to be affected by classical BSSs, which instead
would occupy a region which detaches from the cluster MS at B $\approx$21.5, 
(B-R) $\approx$0.35,
following the extension of the cluster MS below the TO.\\
\noindent
The two classical explanations for these stars are that they are binary
stars or more massive stars born in a separate later
star formation episode.
Our simulations include both these effects in a way that it is
impossible to clarify whether BSSs are present or not.\\
We have visually compared the CMD of NGC~2154 with the CMD of the
rich star cluster NGC~2173 by Bertelli et
al. (2003). The main motivation for this choice is that this study
investigates a cluster -NGC~2173- which is coeval to
NGC~2154 and, interestingly, the authors find that only
a sizable age dispersion can explain the shape of the TO.\\ 
As in our case,
the cleaned CMD of this cluster (their Figs 4 and 5) show only a few
stars on the blue side of the MS, and the region right above the TO is more 
easily explained in terms of binary stars and extended star formation.\\

\noindent
The fact that another coeval cluster -NGC~2173-
does show the same features as NGC~2154 and can be only 
interpreted as having experienced long-lasting star formation 
might instead  be telling us that
the physics of TO mass stars typical of this age might have some
problems, as mentioned above.}

\section*{Acknowledgments}

G. Baume acknowledges financial support from the Chilean {\sl Centro
de Astrof\'\i sica} FONDAP No. 15010003 and from CONICET (PIP~02586).
The work of G. Carraro was supported by {\it Fundaci\'on Andes}.  E.C
and R.A.M. acknowledge support by the Fondo Nacional de
Investigaci\'on Cient\'ifica y Tecnol\'ogica (proyecto No. 1050718,
Fondecyt) and by the Chilean {\sl Centro de Astrof\'\i sica} FONDAP
No. 15010003.

\newpage

\begin{figure*}
\centerline{\psfig{file=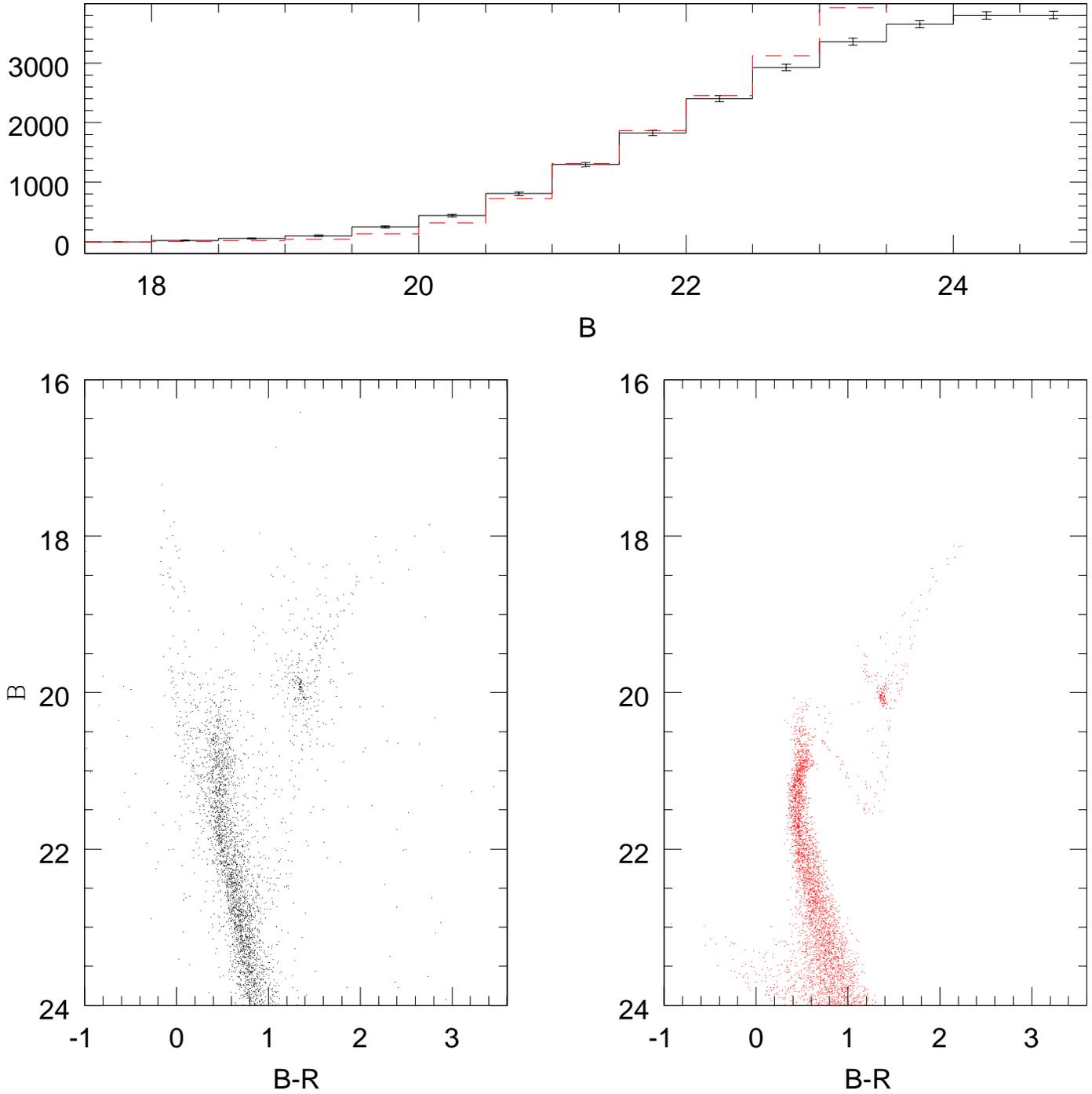}}
\caption{The observational CMD of NGC~2154 is shown in the left panel. The
right panel presents a synthetic CMD of the cluster obtained for our initial
choice of parameters (i.e., assuming $Z=0.005$, an age of 1.7 Gyr, and
30\% of binaries).  The top panel presents the observational ILF
(solid line) together with best model ILF obtained with the above parameters
(dashed line). 1$\sigma$ Poisson error bars are included.}
\end{figure*}

\newpage

\begin{figure*}
\centerline{\psfig{file=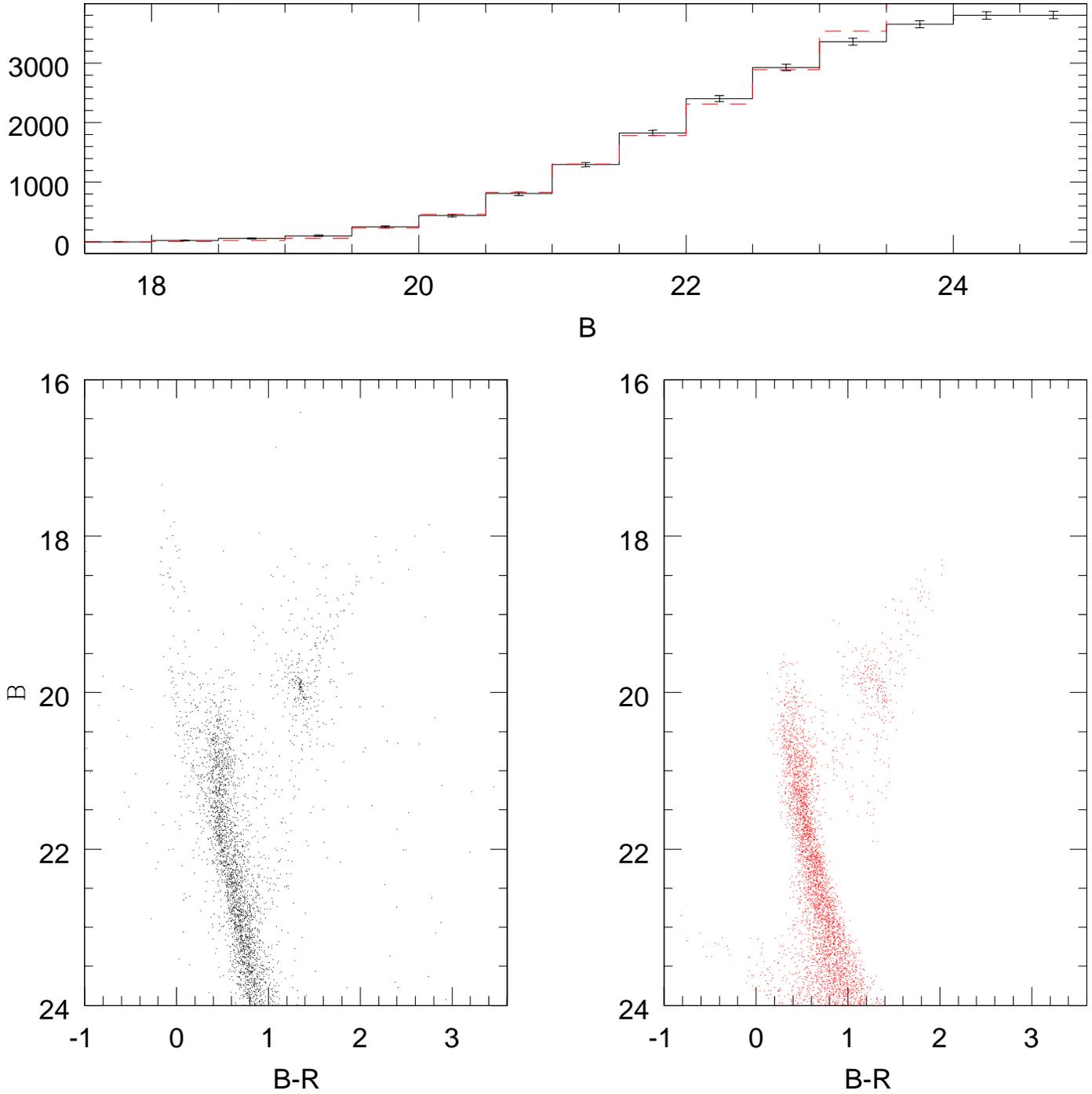}}
\caption{The observational CMD of NGC~2154 is shown in the left panel.
The right panel presents our best fitting synthetic CMD of the cluster.
The top panel presents the observational ILF (solid line) together with
our finally adopted model ILF (dashed line).  1$\sigma$ Poisson error bars
are included. See text for details.}
\end{figure*}

\end{document}